\documentclass[final,5p,times,twocolumn]{elsarticle}

\usepackage{graphicx}

\usepackage{amssymb}

\journal{NIMB}

\begin{document}

\begin{frontmatter}



\title{Altitude distribution of electron concentration in ionospheric D-region in presence of time-varying solar radiation flux }


\author[label1]{A.~Nina}
\ead{sandrast@ipb.ac.rs}
\author[label2]{V.~\v{C}ade\v{z}}
\author[label1]{V.~Sre\'{c}kovi\'{c}}
\author[label3]{D.~\v{S}uli\'{c}}

\address[label1]{Institute of
Physics, University of Belgrade, P.O. Box 57,  Belgrade, Serbia}
\address[label2]{Astronomical Observatory, Volgina 7, 11060 Belgrade, Serbia}
\address[label3]{Faculty of Ecology and Environmental Protection, Union - Nikola Tesla University, Cara Du\v{s}ana 62, 11000 Belgrade,
Serbia}

\begin{abstract}
In this paper, we study the influence of solar flares on electron
concentration in the terrestrial ionospheric D-region by analyzing
the amplitude and phase time variations of very low frequency (VLF)
radio waves emitted by DHO transmitter (Germany) and recorded by the
AWESOME receiver in Belgrade (Serbia) in real time. The rise of
photo-ionization rate in the ionospheric D-region is a typical
consequence of solar flare activity as recorded by GOES-15 satellite
for the event on March 24, 2011 between 12:01 UT and 12:11 UT. At
altitudes around 70 km, the photo-ionization and recombination are
the dominant electron gain and electron loss processes,
respectively. We analyze the relative contribution of each of these
two processes in the resulting electron concentration variation in
perturbed ionosphere.
\end{abstract}

\end{frontmatter}

\section{Introduction}
\label{intr} Characteristics of the ionosphere and their changes are
very important for life and human activity on the Earth. There are
numerous studies about influences of ionospheric disturbances on
operation of powerful energetic systems, navigation and remote radio
communication systems, the atmospheric weather, the human health and
the state of the entire biosphere \cite{goo05,sta06}. Also, the
earthquakes are recently recognized to induce perturbations in the
lowest ionosphere that are registered as specific subionospheric
VLF/LF transmitter signal anomalies (see \cite{hay10} and references
therein).

Methods of investigation of the ionospheric vertical structure are
diverse depending on the applied measuring technics. At higher
altitudes such as the F region (200-700 km), it is possible to
perform direct measurements by probes and satellites while the lower
ionosphere such as the D region (60-90 km) where the altitude range
is too low for satellites and too high for atmospheric balloons,
requires measurements mostly based on radio wave propagation
techniques. The latter approach is the subject of this paper where
we analyze and use the real time VLF signal recordings in recovering
local plasma conditions in the D region. As the presence of
electrons in the ionospheric D-region strongly affects the VLF radio
wave propagation we present a method for determination of the
electron concentration, its time derivative and vertical gradient
profile in the D region. The method is based on recorded time
variations of a chosen VLF signal when the ionospheric D region is
perturbed by a solar flare. For such a case, we apply the analysis
to two distinct regimes characterized by the electron concentration
growing and decreasing in time that we further relate to the regimes
with a dominant electron gain and loss process respectively. The
analysis of VLF signal in these two regimes yields conclusions on
local plasma electron characteristics, chemical processes and
dynamics in the D region. This method is analogous to the method of
ionograms applied to the much higher F region by means of ionosondes
\cite{rad95}.

As said, the structure and characteristics of the ionosphere are not
constant, they vary in time depending on various external
influences. In the sunlit part of the ionosphere, the most important
affect comes from the solar activity which has been considered in
many papers \cite{tho11,tho04}.

At altitudes about 70 km, the dominant electron gain and electron
loss processes are the photo-ionization and recombination
\cite{zig07}. They are in equilibrium, i.e. mutually balanced, when
the ionosphere is unperturbed. However, during periods of increased
solar radiation, the rates of photo-ionization and recombination
processes change in time and, consequently, so does the electron
concentration.

In this work, we choose a typical X ray solar flare recorded by the
GOES-15 satellite on March 24, 2011 from 12:01 UT to 12:11 UT and we
investigate the resulting perturbations induced in the ionosphere.
Time and altitude dependent electron concentration in the ionosphere
is calculated by using the amplitude and phase variations of VLF
signals emitted by the DHO (23.4 kHz) transmitter (Germany) that
were registered by the Belgrade AWESOME receiver system (as a part
of Stanford/AWESOME Collaboration for Global VLF Research) and
numerically processed by the LWPC computing program \cite{fer98} as
done in \cite{gru08,sul10}.

\section{Experimental data and calculation procedure}
\label{exsp}

\begin{figure}
\begin{center}
\includegraphics[width=2.7in]{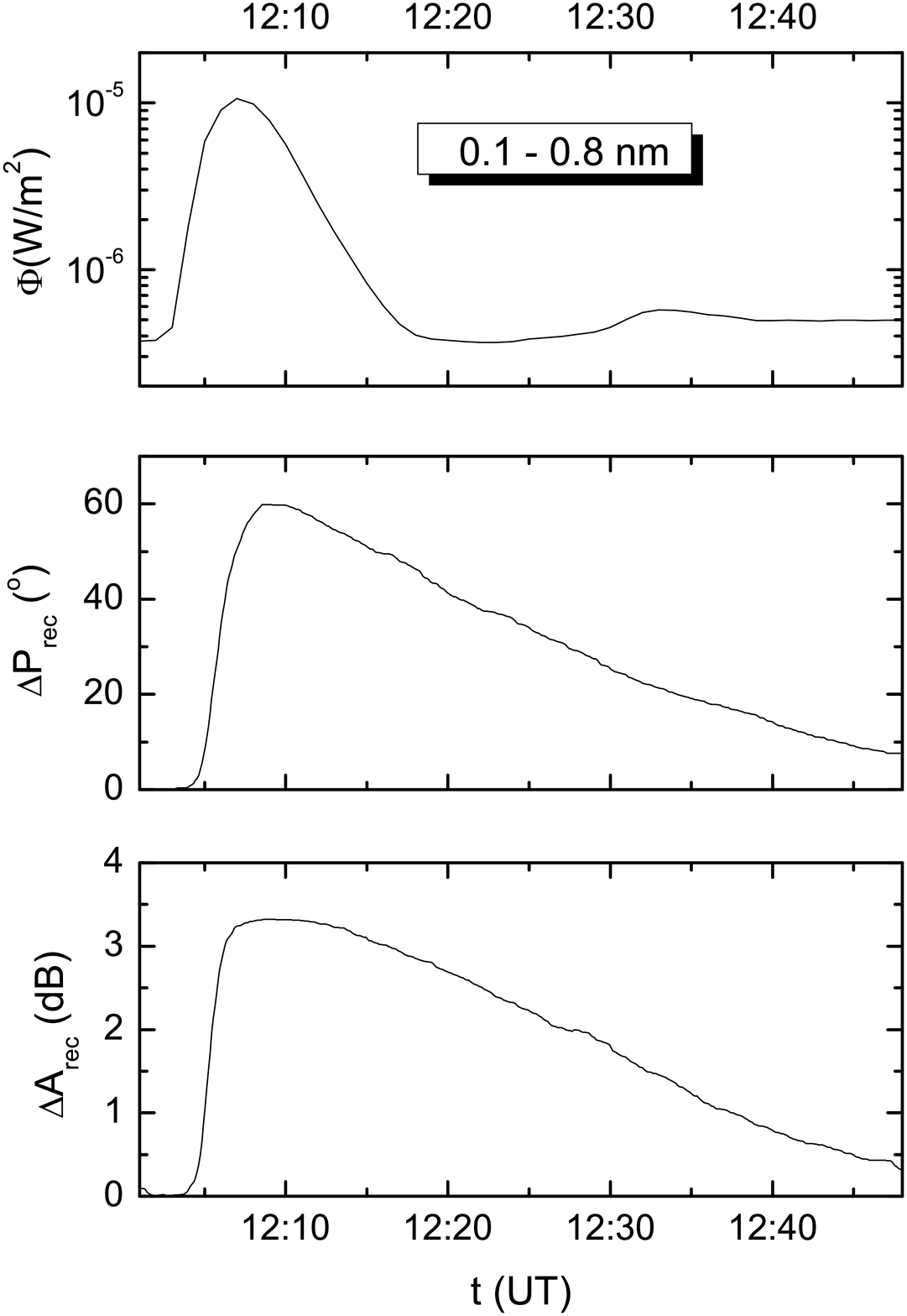}
\caption{Solar flux registered by GOES-15 satellite and phase and
amplitude changes of signal emitted from DHO transmitter (Germany)
and recorded on AWESOME receiver in Belgrade (Serbia) during
observed flares. Zero values correspond to amplitude and phase
recorded when ionosphere is non-perturbed.} \label{f1}
\end{center}
\end{figure}

As of 12:03 UT, March 24, 2011, the Belgrade AWESOME VLF receiver
recorded transient amplitude and phase increases for a signal at
23.4 kHz. Evidently, this was the consequence of a solar X ray
flare, class M1.0, registered by the GOES-15 satellite as can be
seen in Fig.~\ref{f1} showing time dependencies of solar flux (top
panel) at wavelengths between 0.1 and 0.8 nm, and phase and
amplitude time variations of the DHO signal (middle and bottom panel
respectively). The measurable features of the VLF signal refer to
changes in amplitude $\Delta$A$_{\rm rec}$ and phase $\Delta$P$_{\rm
rec}$ expressed in dB and degrees (both taken relative to the
initial unperturbed state) respectively.

Our analysis of electron concentration is based on Wait's model of
the ionosphere \cite{wai64} which is characterized by two
parameters: the wave reflection height  H$^\prime$(t) and sharpness
$\beta$(t) used as input parameters for the LWPC computational
procedure. The resulting computed changes of the VLF signal
amplitude and phase are then compared with corresponding values of
the registered wave which finally yields time dependent values for
H$^\prime$(t) and $\beta$(t) with resolution of 0.1 km and 0.001
km$^{-1}$, respectively.

The electron concentration N(t,h) at time t and altitude h is
calculated using the equation (from \cite{wai64}):
\begin{equation}
N(t,h) = 1.43\times
10^{13}e^{-\beta(t)H^\prime(t)}e^{(\beta(t)-0.15)h}, \label{eq1}
\end{equation}
where the following units are assumed: N(t,h) in m$^{-3}$,
$\beta$(t) in km$^{-1}$, and H$^\prime$(t) and h in km.

The above Eq. (\ref{eq1}) exhibits the physical nature of the
parameter $\beta$(t) as a quantity related to some characteristic
length L(t) defined by L(t)=1/($\beta$(t)-0.15) km which is a
typical e-folding distance for the electron spatial concentration
decrease.

We are now interested in electron concentration in the D-region at
the altitude of around 70 km. Time intervals with dominant
photo-ionization and recombination processes are called the
photo-ionization and recombination regime respectively.

\section{Results and discussion}
\label{res}

Reflections of VLF waves occur at different heights H$^\prime$
depending on electron concentration. When a typical solar flare
occurs, the total emitted radiation first increases in intensity and
then falls off in the after-flare-maximum regime as given in Fig.~1,
top panel. This causes corresponding changes in rates of
photo-ionization and recombination processes which makes the
electron concentration profile time dependent meaning that also the
related VLF wave reflection height H$^\prime$=H$^\prime$(t) and
electron concentration gradient parameter $\beta$=$\beta$(t) become
functions of time t. Fig.~2 shows typical profiles H$^\prime$(t) and
$\beta$(t) resulting from the time varying radiation given in
Fig.~1, top panel. We see that the wave reflecting height
H$^\prime$(t) first decreases with time, reaches a minimum and
starts rising afterwards. This can be explained by the fact that
rise of radiation intensity increases the electron concentration at
all heights which further results in steepening of the electron
concentration gradient, i.e. growth of the parameter $\beta$, and
lowering of the height H$^\prime$ at which the given frequency of
the recorded VLF radio-wave becomes comparable with the local plasma
frequency which itself is a function of electron concentration.
Thus, an incident time-dependent radiation profile with a maximum in
its intensity as given in Fig.~1 (top) causes a functional
time-dependence $\beta$(t) with a maximum and functional dependence
H$^\prime$(t) with a minimum where both extrema occur at the same
time t$_{\rm m}$ as seen in Fig.~2 and they are in a good agreement
with values obtained by \cite{gru08,tho05}. According to Eq.
(\ref{eq1}), two distinct time intervals t$\le$t$_{\rm m}$ and
t$\ge$t$_{\rm m}$ are related to the period of the electron
concentration growth, and to that of electron concentration decrease
respectively (See Fig.~4). These two time domains are also
characterized by the type of the dominant process involving
electrons: the photo-ionization in the first domain and
recombination processes (electron-ion, ion-ion and three body
recombination) in the latter.

The mutual relation of ionospheric parameters $\beta(t)$ and
H$^\prime$(t) is given in Fig.~\ref{f3} for two time intervals
related to the photo-ionization and recombination regime
respectively. The starting point H$^\prime$(t)=74 km and
$\beta$(t)=0.3 km$^{-1}$ is related to the unperturbed ionosphere
before the onset of the considered flare. We see that the curves for
the photo-ionization and recombination regime have different
profiles meaning that $\beta$(t) and H$^\prime$(t) follow different
patterns during the two regimes, i.e. they look like hysteresis
curves between the starting and end position. In principle, the
curve in Fig.~\ref{f3} need not be closed as the starting and the
end point in the $\beta$-H$^\prime$ diagram may be related to
different unperturbed states. Namely, although the ionosphere is
considered unperturbed before the flare onset and after a
sufficiently long relaxation time after the disappearance of the
flare, these two unperturbed states may not coincide as the
unperturbed ionosphere itself is not necessarily stationary and it
may have changed within this time interval for various reasons (like
the Earth rotation, etc.). According to Fig.~\ref{f3}, it can be
concluded that, at a given height H$^\prime$(t) in the D-region, the
parameter $\beta(t)$ is larger during the photo-ionization regime
than during the recombination regime later on.

The resulting affect of photo-ionization and recombination processes
on electron concentration is shown in Fig.~\ref{f4} for different
times during both regimes. We can see that the influence of solar
flare is more pronounced at higher altitudes. At all altitudes, the
electron concentration increases in time in the photo-ionization
regime while it decreases in the recombination regime. Also, the
latter regime lasts longer than the former one.

At higher altitudes, the solar radiation affects more the time and
spatial (with respect to h) derivatives of electron concentration as
seen in Figs.~\ref{f5} and \ref{f6} respectively. In the
photo-ionization regime, the time derivative is positive, it gets
smaller and tends to zero near the electron concentration maximum.
The negative values are obviously related to the recombination
regime when the electron concentration decreases in time tending to
become (quasi)stationary. As to the spatial derivative of the
electron concentration, the photo-ionization and the recombination
regimes exhibit opposite time behaviors: rising and falling in time
respectively at all heights h.

\section{Conclusion}
\label{concl}

In this paper, we have presented an analysis on electron
concentration variation in the ionospheric D-region during the solar
flare of class M1.0 (I$_{\rm max}$(t=12:07 UT)=1.06$\times$10$^{-5}$
W/m$^2$) that occurred on March 24, 2011. Two ionospheric
parameters, the wave reflection high H$^\prime$(t) and sharpness
$\beta$(t), are obtained by comparing amplitude and phase changes of
the signal emitting from the DHO transmitter and recorded by
Belgrade receiver station, with correspondent values calculated by
the LWPC program. From these parameters, the spatial distribution of
electron concentration and its time and height derivatives are
calculated.

We have shown that two characteristic regimes show up: the
photo-ionization regime where the photo-ionization process dominate,
and recombination regime with dominant recombination processes.

In the lower ionosphere, the electron concentration and its spatial
gradient increase with height during both regimes. The electron
concentration thus starts growing due to photo-ionization processes
that prevail after the flare onset, then it reaches a maximum and,
soon after the flare maximum, begins falling off due to the
prevailing recombination processes.

The spatial derivative of the electron concentration shows to be
positive always, e.g. the considered concentration is increasing
function of height h. The reflection height H$^\prime$ is shifting
downwards during the photo-ionization regime and upwards during the
recombination regime.

As seen in Fig.~\ref{f5}, the time derivative of the electron
concentration is positive and negative in the photo-ionization and
recombination process respectively.

Finally, we see that the solar flare impact on the shape of the
electron concentration profile grows with height.

\section*{Acknowledgment}
The present work was supported by the Ministry of Education and
Science of the Republic of Serbia as a part of the projects no.  III
44002, 176002 and 176004.

\clearpage
\begin{figure}
\begin{center}
\includegraphics[width=2.7in]{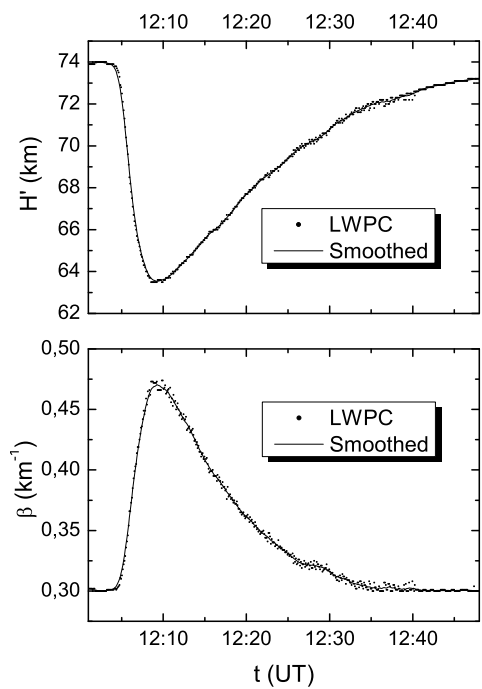}
 \caption{The reflection hight H$^\prime$(t) and sharpness $\beta$(t)
 obtained by comparative LWPC simulation and
recorded signal characteristic values.} \label{f2}
\end{center}
\end{figure}
\begin{figure}
\begin{center}
\includegraphics[width=3.1in]{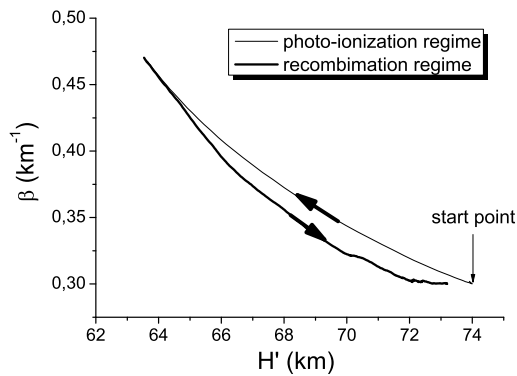}
\caption{The connection between ionospheric parameters H$^\prime$(t)
and $\beta$(t) during photo-ionization and recombination regimes. }
\label{f3}
\end{center}
\end{figure}
\begin{figure}[h]
\begin{center}
\includegraphics[width=3.8in]{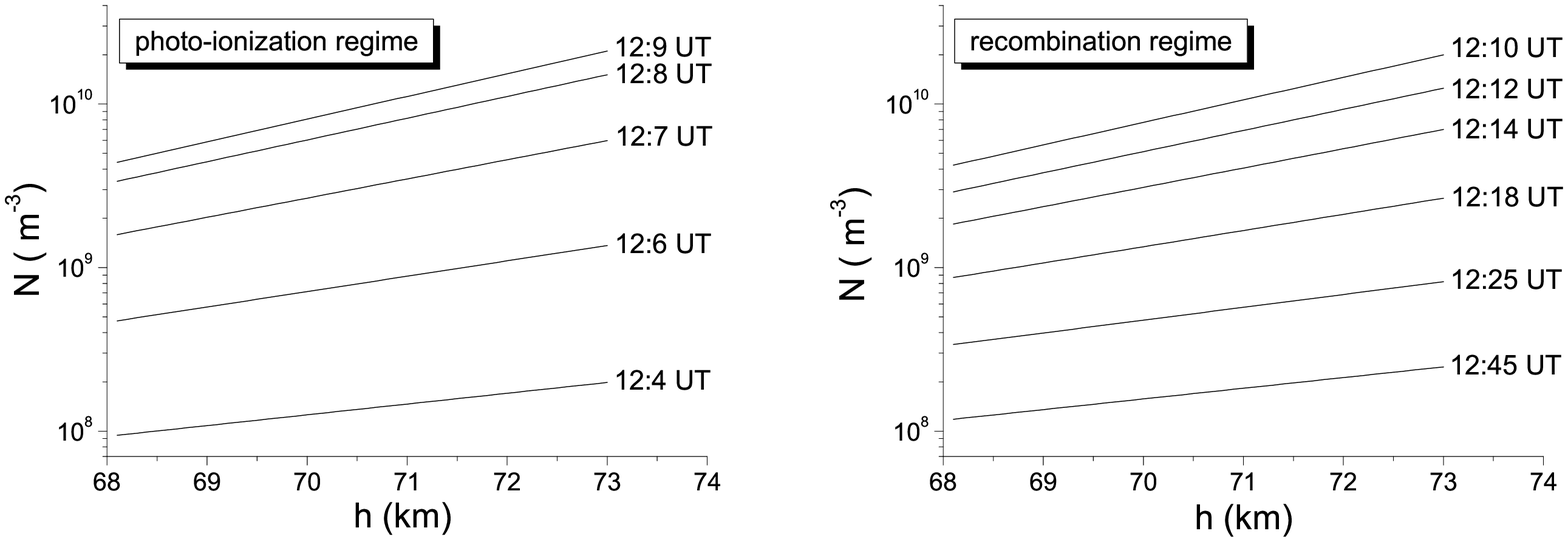}
\caption{The vertical distribution of electron concentration during
photo-ionization and recombination regimes.} \label{f4}
\end{center}
\end{figure}
\begin{figure}[h]
\begin{center}
\includegraphics[width=3.8in]{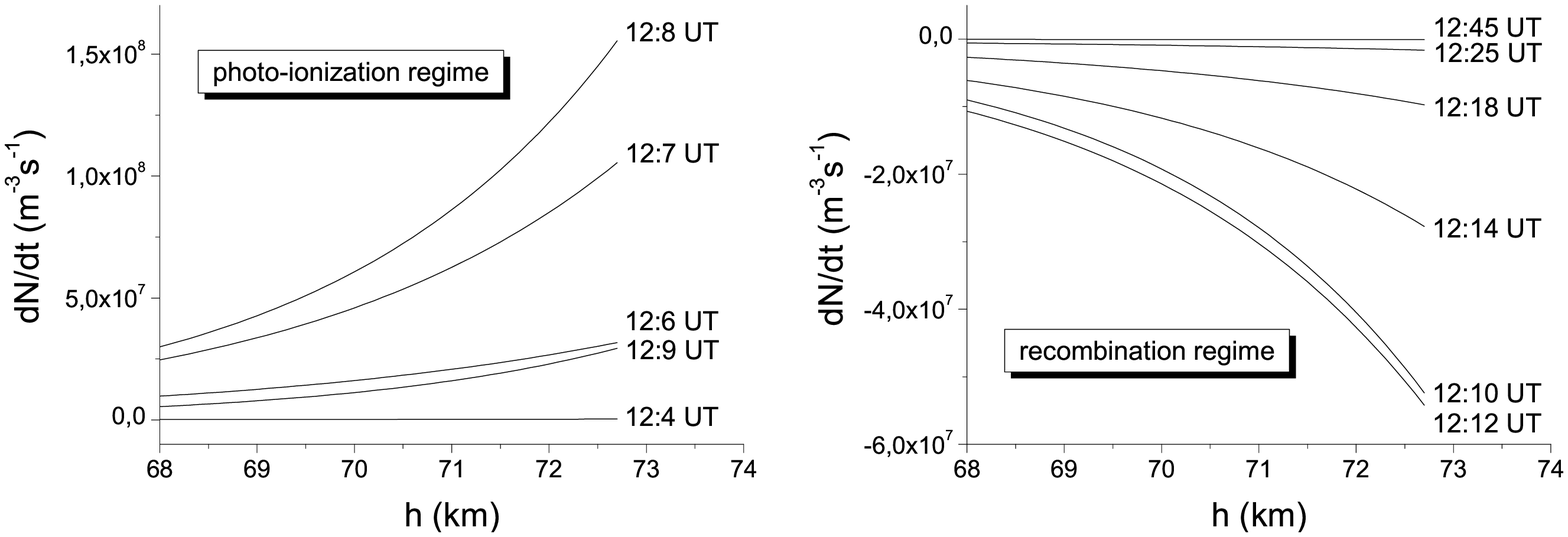}
\caption{The vertical distribution of t-derivative of electron
concentration during photo-ionization and recombination regimes.}
\label{f5}
\end{center}
\end{figure}
\begin{figure}
\begin{center}
\includegraphics[width=3.8in]{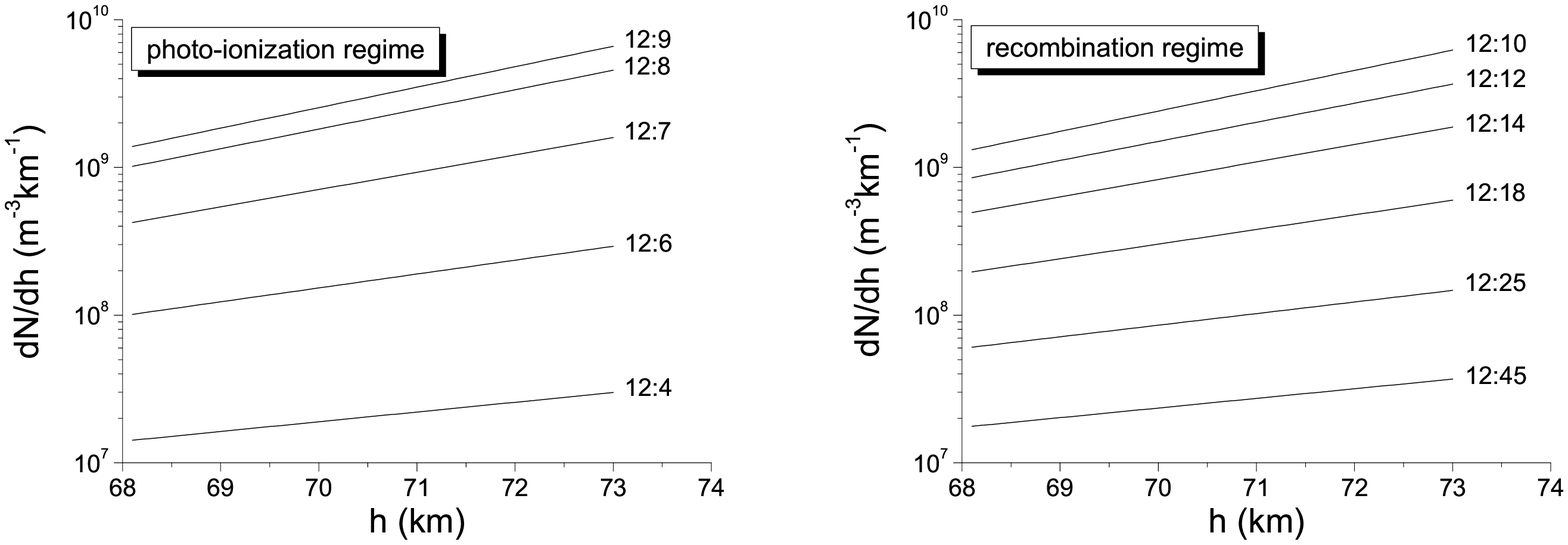}
\caption{The vertical distribution of  h-derivative of electron
concentration during photo-ionization and recombination regimes.}
\label{f6}
\end{center}
\end{figure}

\end{document}